\documentstyle[11pt]{article}

\newcommand{\br}{{\bf r}}
\newcommand{\ba}{{\bf a}}

\newcommand{\da}{{\cal {A}}}
\newcommand{\ca}{{\bf{\cal {A}}}}
\newcommand{\dda}{{\cal C}}

\begin{document}

\font\ninerm = cmr9

\baselineskip 14pt plus .5pt minus .5pt

\def\footnoterule{\kern-3pt \hrule width \hsize \kern2.6pt}

\hsize=6.0truein
\vsize=9.0truein
\textheight 8.5truein
\textwidth 5.5truein
\voffset=-.4in
\hoffset=-.4in

\pagestyle{empty}
\begin{center}
{\large\bf Moduli Space Dynamics  of a First-Order 
Vortex System}\footnote{\ninerm
\hsize=6.0truein This work was
supported in part by funds provided by 
the Korea Science and Engineering Foundation through
the SRC program of SNU-CTP, and
the Basic Science  Research Program
under project \#BSRI-96-2425.}
\end{center}

\vskip 1.5cm

\begin{center}
{\bf Dongsu Bak\footnote{\ninerm
\hsize=6.0truein email address: dsbak@mach.scu.ac.kr}}\\
{\it Department of Physics,
Seoul City University,
Seoul 130-743, Korea}
\end{center}
\begin{center}
{\bf Hyuk-jae Lee\footnote{\ninerm
\hsize=6.0truein email address: hjlee@theory.yonsei.ac.kr}}\\
{\it Institute of Natural Science, Yonsei University, 
Seoul 120-749, Korea}
\end{center}

\vspace{1.2cm}
\begin{center}
{\bf ABSTRACT}
\end{center}
The moduli space dynamics of  vortices 
in 
the Jackiw-Pi model where a non-relativistic Schr\"odinger 
field couples minimally 
to Chern-Simons 
gauge field, is considered. It is shown that the difficulties 
in direct
application of Manton's method to obtain a moduli-space metric 
in the first order
system can be
circumvented by turning the Lagrangian into a second order 
system. 
We obtain exact metrics for some simple cases and describe   
how the vortices 
respond to an  external U(1) field. We then construct an 
effective Lagrangian
describing dynamics of the vortices. In addition, we clarify  
strong-weak coupling duality  
between
fundamental particles and vortices.

\bigskip
\begin{center}
(PACS: 11.15.Kc, 11.27.+d, 03.65.Ca)
\end{center}
\bigskip
\begin{center}
Submitted to: PLB 
\end{center}
\bigskip
\bigskip

\vfill
\space\space SCU-TP-97-1003 \space\space
 \space\space
SNUTP-97-077 
\hfill
June 1997
\eject

\baselineskip=16pt plus 3pt minus 3pt

\pagestyle{plain}
\pagenumbering{arabic}
\setcounter{page}{1}

\pagebreak[3]
\setcounter{equation}{0}
\nopagebreak
\medskip
\nopagebreak


Vortices play an important role in
describing some 2+1 dimensional systems. Most of them arise 
as solitonic
configurations in field-theoretic 
descriptions\cite{olesen,ykim,jackiw}, 
as in the case of
nonabelian magnetic monopoles in 3+1 dimensions. 
There are limited 
number of 
exact multi-vortex solutions and this partly explains why 
descriptions of their dynamics are mostly 
involved\cite{ruback}.
When a theory admits static multi-vortex solutions 
saturating Bogomol'nyi 
bound\cite{bogomolnyi}, the 
low energy dynamics of slowly moving vortices is well 
approximated  by
geodesic motions in the moduli manifold, which may be 
explained as follows. 
As the moduli space consists of 
a collection
of points which give the same value of energy functional, the slow 
motions in moduli
space can be  effectively described by the kinetic terms 
produced by the 
time dependence of  moduli parameters.  This is Manton's idea that 
was successfully 
applied 
to the case of  BPS magnetic monopoles\cite{manton}.

Exact multi-vortex solutions were found in Jackiw-Pi model where 
a non-relativistic
Schr\"odinger field  minimally couples to the Chern-Simons gauge 
potentials\cite{jackiw}. 
Previous attempts
to analyze dynamics of the vortices within moduli space 
approximation 
 are incomplete 
mainly because
the model is first order in time derivatives and hence the metric 
does not appear by
direct application of the Manton's methods\cite{hua}. This problem, 
of course, 
persists in all first-order system such as
non-relativistic models and relativistic fermionic models.   
In Ref.~\cite{liu}, 
an extra phase
to the static Schr\"odinger field solution  was introduced 
to overcome the 
difficulty and it was 
shown that the
leading order kinetic term can be properly produced. But the  role of the 
phase was unclear and, 
moreover, it was not fixed  by 
static configuration of vortices  but rather trial 
functions were used 
in their framework.     

In this note, we observe a simple fact that the Lagrangian 
can be turned into
second-order form in time by eliminating momentum variables.
It will be shown that  the phase in Ref.~\cite{liu} is 
indeed determined by
the continuity equation  in terms of solely vortex configurations.
Once a second-order
formulation is achieved, one may apply the standard Manton's method
of evaluating the kinetic term in term of 
moduli parameters  
to derive the
moduli space metric. 

Equipped with the second-order Lagrangian, we shall obtain 
exact metrics in the cases of a vortex and coincident 
multi-vortices. 
We also study how the vortices respond to an external U(1)
field. 
We then describe the structures and asymptotic behaviors 
of the metric 
for  spatially separated vortices.
Based on these analyses, we will construct an effective 
Lagrangian describing
the dynamics of well-separated vortices.   

Finally, we observe that there is an intriguing dual 
structure between 
fundamental
particles and the vortices described by the effective Lagrangian. 
Namely, at weak (strong) 
statistical interactions between the particles, the vortices 
feel strong (weak) 
statistical interactions.

The non-relativistic Schr\"odinger field minimally coupled to 
the Chern-Simons gauge field is described by the Lagrange density  
\begin{equation} 
{\cal L}={\frac{{\kappa }}{2}}\epsilon ^{\alpha \beta \gamma 
}A_{\alpha }\partial_\beta A_\gamma +i\psi D_{t}\psi -
{\frac{1}{2m}}|{\bf D}\psi 
|^{2}+{{g}}(\psi ^{*}\psi )^{2},  \label{lag} 
\end{equation} 
where $D_{t}=\partial _{t}+iA^{0},{\bf D}=\nabla -i{\bf A}$ 
and $\kappa$ is taken to be positive for simplicity.  
By solving the Gauss-law constraint
\begin{equation}
\label{gauss} 
\epsilon^{ij}\partial_i{A}^j-{\rho\over \kappa}=0, \ \ \ 
(\rho\equiv |\psi|^2) 
\end{equation} 
the system may be equivalently described by
\begin{equation} 
{\cal L}=i\psi \partial_{t}\psi -{\frac{1}{2m}}|(D_x-iD_y)\psi|^{2}
+({{g}}-{1\over 2m\kappa})(\psi ^{*}\psi )^{2},  \label{lagb} 
\end{equation} 
where 
\begin{equation} 
{\bf A}(t,{\bf r})=\frac{1}{\kappa }\int d{\bf r}^{\prime }
{\bf G}({\bf r%
}-{\bf r}^{\prime })\rho (t,{\bf r}^{\prime })  \label{gauge} 
\end{equation} 
with ${\bf G}$ being the Green's function  
\begin{equation} 
G^{i}({\bf r})=\frac{1}{2\pi }\partial_{i}\theta (\br)={\frac{{%
\epsilon ^{ij} r^{j}}}{{2\pi r^2}}} .  \label{green} 
\end{equation} 
Quantization of this model leads to $N$-body Schr\"odinger 
equation describing 
particles interacting via Aharonov-Bohm potential\cite{bak1}. 
The coupling strength
of the interaction is characterized by {\it statistics parameter} 
$\nu\equiv 1/(4\pi\kappa)$,
$2\pi$ multiple of which is the phase acquired when two particles 
are exchanged. 
Other than obvious gauge and Galilean symmetries, this system
possesses SO(2,1) group invariance  comprizing usual 
time translation, 
time dilation
$T(t)=at$, and special conformal transformation $T(t)= t/(1-at)$. 
This conformal symmetry
is in general broken by quantum corrections,  but there is a 
critical strength  
of $g={1/(2m\kappa)}$ where the symmetry persists even at 
the quantum 
level\cite{bergman}. 
At this critical strength, the system also admits static 
multi-vortex
solutions saturating the energy bound characterized by the self-dual  
 equation,
\begin{equation} 
\label{bound}
(D_x-iD_y)\psi =0
\end{equation} 
In what follows we shall fix the value of $g$  to be critical.  
The general solutions of the self-dual equation were found in 
Ref.~\cite{jackiw} 
and given by 
\begin{equation} 
\psi(z)={\frac{{2\kappa^{1\over 2} |f^{\prime}(z)|}}
{{(1+|f(z)|^2)}}} e^{i\omega(z)}, \qquad 
f(z)=\sum^N_{n=1} {\frac{{c_n}}{{z-z_n}}}, 
\label{sol} 
\end{equation} 
with 
\begin{equation}  
\omega = Arg(f'V^2),\qquad V(z)=\prod^N_{n=1}(z-z_n)  \label{sol1} 
\end{equation}   
where $z = x +i y$ and $z_n(=a^x_n+ia^y_n)$ are complex 
constants. 
It describes $N$ separated solitons at positions $\ba_n$ 
with scales 
and phases $c_n$. This accounts for the fact that the dimension 
of $N$-vortex 
modular space 
is $4N-1$ upon elimination of one overall phase parameter.

As stated earlier, the action may be turned into a second-order 
form  by eliminating
momentum variables. First, we rewrite the action (\ref{lagb}) 
in terms of $\rho(\br,t)$ and 
$\Theta(\br,t)$ with  $\psi=\rho^{{1\over 2}}e^{i(\Theta+\omega)}$, 
where $\omega$
is a fixed background phase function. Later this will be identified 
with the phase of
the static soliton (\ref{sol}). 
In  terms of new variables, the transformed 
Lagrange density reads 
\begin{equation} 
{\cal L}={ \Theta} \dot{\rho} - \rho\dot{\omega} -
{1\over m}\{\frac{1}{8\rho} 
\nabla\rho\cdot\nabla\rho +{\rho\over 2}({\bf \da}-\nabla\Theta)^2-
{\rho^2\over 2\kappa}\}. 
\label{lagc} 
\end{equation} 
where ${\bf\da}$ being ${\bf A}-\nabla\omega$. 
The $\Theta$ variation of the above Lagrange density gives rise 
to the constraint on $\Theta$,
\begin{equation} 
\dot{\rho}  
=-{1\over m}\nabla\cdot \rho ({\ca}-\nabla{\Theta}). 
\label{eom} 
\end{equation} 
In fact, this  is nothing but the continuity equation for 
 charge. Moreover, we clearly see that the 
phase introduced in Ref.~\cite{liu} 
can
now be determined from the vortex configurations.  
We solve the constraint with respect to $\Theta$
\begin{equation} 
{\Theta}(\br)= \int d{\bf r}' K^{-1}({\bf r},{\bf r}')
(m\dot{\rho}({\bf r}')
-\nabla \cdot \rho\ca (\br')), 
\label{theta} 
\end{equation} 
where we define  
\begin{eqnarray} 
&&K\equiv -\nabla\cdot\rho\nabla \\ 
&&\int d\br' K^{-1}({\bf r},{\bf r}')K({\bf r}',{\bf r}'') 
=\delta^2({\bf r}-{\bf r}'').  
\label{ker} 
\end{eqnarray} 
Inserting (\ref{theta}) into (\ref{lagc}) and dropping 
irrelevant total
derivative terms, we obtain the desired second-order 
Lagrangian:
\begin{eqnarray} 
{L}&=& \frac{m}{2}\int d 
{\bf r} d{\bf r}'\{\dot{ 
\rho}({\bf r})\!-{1\over m}\nabla\cdot \rho\ca\} K^{-1}({\bf r},
{\bf r}')\{\dot{\rho}({\bf r}')
\!-{1\over m}
\nabla\cdot \rho\ca\}
-\int \!d{\bf r}\rho \dot{\omega}\nonumber\\
&-&\int \! d\br {1\over 2m}
\left(\frac{1}{4\rho} 
\nabla\rho\!\cdot\!\nabla\rho\! +\rho\da^2\!-\!
{\rho^2\over \kappa}\right), 
\label{lagd} 
\end{eqnarray}  
The redundant introduction of background phase $\omega$ 
in the derivation
of (\ref{lagd}) can be justified if one considers the 
vortex configurations in (\ref{sol}). Namely, 
$\nabla^2 \omega_{\rm sol}$ with 
the phase $\omega_{\rm sol}$ in (\ref{sol1}) contains  
delta-function 
contributions  and their separation by the background phase gives  
some convenience in performing integration by parts with 
vortex wave functions. 

Following Manton's idea, one substitutes the multi-vortex 
solution (\ref{sol})
with time dependent moduli parameters $\xi^i$ into 
(\ref{lagd}) to obtain 
$N$-soliton effective Lagrangian,
\begin{eqnarray} 
{L_{\rm eff}}&=&  \frac{m}{2}\int d 
{\bf r} d{\bf r}'\dot{ 
\rho}({\bf r};\xi) K^{-1}({\bf r},{\bf r}')\dot{\rho}({\bf r}';\xi)
-\int d{\bf r}\rho(\br;\xi) \dot{\omega}(\br;\xi) \nonumber\\
&=&\sum_{i,j}g_{ij}({\bf\xi})\dot\xi^i\dot\xi^j + 
\sum_i C_i({\bf\xi})\dot\xi^i,  
\label{efflag} 
\end{eqnarray} 
where we have used the fact that $\nabla\cdot \rho\ca$ 
and the last integral 
in (\ref{lagd}) are vanishing when evaluated upon the 
vortex solution. 
    The generic 
expression of $C_i({\bf\xi})$ for $N$ soliton solutions 
was given in 
Ref.~\cite{hua,liu}. Hence we shall mainly focus on 
finding the metric.

Specialize in the N-coincident vortices described by
\begin{equation} 
\psi(r)=\frac{2\kappa^{{1\over 2}} N}{|\br-\ba|}
\left(\frac{c^N}{|\br-\ba|^N}  
+\frac{|\br-\ba|^N}{c^N}\right)^{-1}e^{i(1-N)\theta}, 
\label{nsol} 
\end{equation} 
where  $\ba$ is the position of the vortex  
and the remaining positive quantity $c$ is related to 
the dilatation of  vortices. 
Of course these three degrees of freedom do not account 
for the full moduli space
but rather one is specialized in coherent motions and 
dilatation of $N$ vortices. 
We then proceed by evaluating first 
\begin{equation} 
{U}(\br;\xi,\dot\xi)= \int d{\bf r}' K^{-1}({\bf r},{\bf r}')
\dot{\rho}({\bf r}';\xi), 
\label{atheta} 
\end{equation} 
by solving the differential equation
\begin{equation} 
\nabla\cdot\rho\nabla U=-\dot{\rho}  . 
\label{eqtheta} 
\end{equation} 
Though the above equation resembles the two-dimensional 
Laplace equation 
with nonflat metric, the method of extracting solution 
is not known for 
generic $\rho$. However, with 
the specific choice of $\rho$ in (\ref{nsol}), one easily finds   
\begin{equation} 
{U}=({\bf r}-\ba)\cdot \dot{\bf a}  
+\frac{\dot{c}}{2c}|{\bf r} -{\bf a}|^2. 
\label{expu} 
\end{equation} 
Inserting the expression $U$ into (\ref{lagd}) and 
performing integration, one 
is led to the effective Lagrangian
\begin{equation} 
\label{neff}
{ L}_{\rm eff}=\frac{1}{2} M \dot{\bf a}\cdot \dot
{\bf a} + \dot{c}\dot{c} 
{K(c)\over c^2},
\end{equation} 
where $M=4\pi\kappa mN$    
and the special conformal charge $K$ is given by 
\begin{equation}  
K(c)\equiv {m\over 2}\!\int d\br \rho(\br-\ba;c) |\br-\ba|^2 =
2\pi m\kappa c^2 {{\pi}/{N} \over {\sin {\pi}/{N}}}.
\end{equation} 
One sees that the vortex mass is proportional to its number and 
 no  statistical interaction arises for the overall translation. 
For $N\neq 1$, energy cost required for 
the finite velocity excitation in the conformal
mode  is of the same order as the  spatial 
velocity. In case $N=1$,
the charge diverges and the excitation in this mode 
is forbidden due to 
its infinite energy cost.
In generic 
configuration, one 
may show that the inertia for 
this overall
conformal mode is measured by $K(c)/c^2$. Noting that 
$K$ in general 
diverges unless $\sum_{m=1}^N
c_m=0$~\cite{jackiw}, one conclude that the allowed conformal 
mode in this coincident vortex 
is rather exceptional.

Motion of two vortices in the moduli space is in general 
 complicated
 especially when they approach closely. For simplicity, 
we consider 
here a configuration characterized by a special choice of $f(z)$
\begin{equation} 
f(z)={1\over 2}\left({c\over z-z_a}-{c\over z+z_a}\right) ,
\label{ftwosol} 
\end{equation} 
where the complex variable $z_a$ denotes $a_x +ia_y$. 
The static solution with this choice is given as
\begin{equation} 
\psi(r)={ 4\kappa^{{1\over 2}}c  z_a z \over c^2a^2 
+|z^2-z_a^2|^2}, 
\label{twosol} 
\end{equation}
Here the center of mass frame is chosen  to focus on 
the relative motion.  
The two solitons have the same size and the relative 
phase between $c_1$ 
and $c_2$ is fixed to the above specific value.  
From the symmetry and scaling argument, one may easily 
show that the metric has 
the following form in the  effective Lagrangian,
\begin{equation} 
\label{twoeff}
{L}_{\rm eff}=g_{aa}(b) \dot{a}^2 +  g_{\theta\theta}(b) 
a^2 {\dot{\theta}}^2(\ba) 
+g_{cc}(b)\dot{c}^2 +2 g_{ac}(b)\dot{a}\dot{c} -8\pi\kappa 
{{d\over dt}{\theta}}(\ba), 
\end{equation} 
where $b$ is the ratio $c/a$. One find the last statistical 
interaction term
agrees with that in (\ref{neff}) with $N=2$. 
Explicit evaluation of the metric requires solving 
(\ref{eqtheta}) with 
$\rho$ specified by (\ref{twosol}). Due to the difficulty involved 
with the evaluation, we consider here rather restricted cases.
We first consider the limit $c\rightarrow 0$, which 
describes two solitons 
at large separation compared to their size. 
 In this limit, the metric ($g_{aa}$,$g_{\theta\theta}$) 
must be constant 
in $a$ since 
they only depend on the vanishing ratio $b$.  In this 
case,  $\rho$ 
takes a simple form,
\begin{equation} 
\rho(\br)={ 4\pi\kappa}(\delta^2(\br-\ba) +\delta^2(\br+\ba)), 
\label{tworho} 
\end{equation}   
which reflects that the profile is concentrated at each location.
With this expression, the equation (\ref{eqtheta}) is solved by
$U={\dot\ba}\cdot \br \,\,{\rm Sign}(\ba\cdot\br)$. 
Evaluation of integrals in (\ref{efflag}) leads  to 
the effective Lagrangian
\begin{equation} 
\label{twoseff}
{ L}_{\rm eff}=4\pi \kappa m\dot{\bf a}\cdot \dot{\bf a}  
-8\pi\kappa 
{d\over dt}{\theta}(\ba). 
\end{equation} 
The other case comes with the restriction of 
motion by the condition ${d\over dt}(a-c)={d\over dt}
\theta(\ba)=0$. Again 
the equation in  (\ref{eqtheta}) can be solved with 
this restricted $\dot{\rho}$ by $U={\dot{c}r^2/ (2c)}$. 
Then the following combination 
of the metric components can be determined:
\begin{equation} 
g(b)\equiv g_{aa}(b)+ 2b g_{ac}(b)+ b^2 g_{\theta\theta}
(b)= 4\pi\kappa m\sqrt{1+b^2}
{ E}\left({1\over\sqrt{1+b^2}}\right)
\label{kmetric} 
\end{equation}    
where $E(x)$ is the complete elliptic integral of the 
second kind. Taking 
the limit $b\rightarrow 0$, one regains $g_{aa}=4\pi\kappa$ 
with help of $E(1)=1$, 
which agrees with 
the metric in (\ref{twoseff}). 

We now turn to the  problem of how vortices respond to 
an external U(1) field. 
We shall assume the external field is sufficiently weak 
so that the vortex 
configurations are not deformed considerably. In this 
weak-field  probe, 
the response of the vortices is linear in 
the applied external field. 
The system is described by the Lagrange density
\begin{equation} 
{\cal L}=i\psi \partial_{t}\psi -A^e_0|\psi|^2 -
{\frac{1}{2m}}|[(D_x-iA_x^e)-i(D_y-iA_y^e)]\psi|^{2},  
\label{lagext} 
\end{equation}  
where the external gauge potential $A^e_\mu$ couples 
minimally to the Schr\"odinger field.
Adopting  similar methods used for the case without 
external field, 
and 
keeping to the linear terms in the external field, one obtains 
an effective 
Lagrangian for a generic soliton solution:
\begin{eqnarray} 
{ L}&=& -\int d{\bf r}\rho(\br) \dot{\omega}(\br) + 
\frac{m}{2}\int d 
{\bf r} d{\bf r}'\dot{ 
\rho}({\bf r}) K^{-1}({\bf r},{\bf r}')\dot{\rho}({\bf r}')   
\nonumber\\
&-&\int d\br\rho (
A_0^e-{\bf A}^e\cdot \nabla U) +{1\over 2m} 
\int d\br \rho\epsilon^{ij}\partial_i A_e^{j},
\label{effextlag} 
\end{eqnarray}
where $U$ is defined by the relation (\ref{atheta}). 

When the solution in (\ref{sol})  is used, the effective 
Lagrangian 
can be easily computed in the limit where each $c_i$ goes 
to zero. The result
is given by
\begin{equation} 
\label{twoexteff}
{ L}_{\rm eff}\!=\!2\pi\kappa m\sum_{n}\dot{\bf a}_n\!\cdot 
\dot{\bf a}_n  
\!-\!4\pi\kappa \sum_{l\neq n}{d\over dt}{\theta}(\ba_l\!-\!\ba_n)
\!-\!4\pi\kappa \sum_{n} \{A^e_0(\ba_n)\!-\!\ba_n\cdot
{\bf A}^e(\ba_n)\!-\!{{B^e(\ba_n)\over 2}}\}
\end{equation} 
where we have used the result (for well-separated vortices) 
given in Ref.~\cite{hua} 
for the statistical  interaction term. 
Each vortex couples to the external fields with a coulping 
strength $4\pi\kappa$ and 
carries a magnetic moment $2\pi\kappa$.   

Based on the above investigation, one may finally deduce 
$N$-soliton
effective  Lagrangian with a help of   Chern-Simons kinetic 
term. We consider
dynamics of well separated vortices, and each vortex 
producing the Aharonov-Bohm 
 potential
as indicated in the statistical interaction terms in 
above effective Lagrangians.
Each vortex is supposed to feel the gauge potential 
produced by other vortices. 
This reasoning
is summarized in the following local effective Lagrangian:
\begin{eqnarray} 
{L_{{\rm sol}}}={\kappa'\over 2}\int\! d\br 
\epsilon^{\alpha \beta \gamma 
}\dda_{\alpha }\partial_\beta \dda_\gamma +\sum_{n}
\{{m'\over 2}\dot{\bf a}_n\cdot 
\dot{\bf a}_n 
-\dda_0(\ba_n)+\dot\ba_n\cdot{\bf\dda}(\ba_n)\}
\label{finallag} 
\end{eqnarray} 
with $\kappa'= 1/(16\pi^2 \kappa)$ and $m'=4\pi\kappa m$.
Here the magnetic moment interaction 
is ignored since the magnetic field produced by vortices 
are localized at 
the vortex locations.

In this effective theory approach,
one finds 
that there is a duality between the soliton 
Lagrangian and the original $N$ particle Lagrangian. Namely, the 
dimensionless coupling constants
$\nu\equiv 1/(4\pi\kappa)$  and $\nu'\equiv 1/(4\pi\kappa')$  
are related by
\begin{equation} 
\label{duality}
\nu\nu'=1, \ \ \ \  {\rm with}\ \ \nu\sim \nu+1, \  
\nu'\sim \nu'+1
\end{equation} 
where the equivalence classes of $\nu$/$\nu'$ in particle/soliton 
sector  come from 
the fact that these shift
do not change statistics nor the physical amplitudes 
within each sector. [For example, the cases $\nu\!=\!1$ 
and $\nu\!=\!2$ 
which are equivalent in particle sector 
correspond to
inequivalent interactions ($\nu'\!=\!1$ and $\nu'\!=\!1/2$) in 
soliton sector.]
This  certainly is a strong and
weak coupling duality since the weak (strong) coupling in 
the particle sector implies
strong (weak) coupling in the soliton sector.

In this note, we have investigated slowly moving vortices 
in the Jackiw-Pi 
model by Manton's method.
One interesting problem we do not deal with  in this note, 
is the problem of head-on
collision of two vortices.  To find their motion, one needs  
a detailed form 
of the metric $g_{aa}$ and $g_{\theta\theta}$ of the 
two solitons.
(In the cases of Abelian-Higgs model, it is shown that 
there are $\pi/2$-angle scattering
in head-on collision of two vortices\cite{ruback}.) 
Admitting the difficulty involved 
in analytic evaluation of the 
two vortex metric,
it may be probable to resort to numerical works. 

Another point stressed 
is that the methods developed in this note might be 
applied to generic
first-order vortex system once there exist static 
multi-vortex solutions. This includes 
the vortices in Jackiw-Pi model with nonvanishing 
chemical potential\cite{harin} 
and the various models 
considered in Ref.~\cite{manton2}. If the method 
turns out to be  
problematic in analyzing these models, the reason or
its possible modification need to be clarified.

 We close this note with a remark that the origin of 
the duality might have to do with
supersymmetry. Supersymmetric version of Jackiw-Pi 
model was studied in Ref.~\cite{min2}. 
Detailed analyses of the soliton sector in this
 supersymmetric model 
will be of great interest.

\bigskip
\begin{center}
{\bf ACKNOWLEDGEMENTS}
\end{center}
\ \indent
The authors would like to  thank R. Jackiw, Y. Kim, K. Lee, 
and H. Min for 
enlightening discussions and J. Lee for 
critical reading of the manuscript.
\hfill
   
\bigskip
\bigskip



\end{document}